\documentclass[twocolumn,showpacs,preprintnumbers,amsmath,amssymb]{revtex4}

\usepackage{graphicx}
\usepackage{dcolumn}
\usepackage{bm}

\begin{document}

\title{Ambipolar Nernst effect in NbSe$_2$}

\author{Romain Bel, Kamran Behnia} \affiliation{Laboratoire de Physique Quantique,
 (CNRS), ESPCI,10 Rue de Vauquelin  F-75231 Paris, France}
\author{Helmut Berger}\affiliation{D\'epartment de Physique, Ecole polytechnique F\'ed\'erale de Lausanne, CH-1015
Lausanne, Switzerland},

\date{February 13, 2003}

\begin{abstract}
The first study of Nernst effect in NbSe$_2$ reveals a large
quasi-particle contribution with a magnitude comparable and a sign
opposite to the vortex signal. Comparing the effect of the Charge
Density Wave(CDW) transition on Hall and Nernst coefficients, we
argue that this large Nernst signal originates from the
thermally-induced counterflow of electrons and holes and indicates
a drastic change in the electron scattering rate in the CDW state.
The results provide new input for the debate on the origin of the
anomalous Nernst signal in high-T$_c$ cuprates.
\end{abstract}

\pacs{72.15.Jf,71.45.Lr,74.70.Ad}

\maketitle

Since the report by Xu \emph{et al.}\cite{xu} on the detection of
a finite Nernst signal in the normal state of underdoped cuprates,
this less common thermoelectric coefficient has become a focus of
renewed attention\cite{wang1,wang2,capan}. A well-established
source of Nernst signal is the movement of vortices induced by a
thermal gradient in the vortex-liquid state of a type II
superconductor\cite{ri}. In metals, on the other hand, the Nernst
coefficient, while much less explored, is believed to be small.
The fundamental reason behind this belief, recently recalled by
Wang \emph{et al.}\cite{wang1} and dubbed Sondheimer cancellation,
was first put forward in 1948\cite{sondheimer}.

In this Letter, we present the case of NbSe$_2$. A large negative
Nernst coefficient, persisting at temperatures well above
T$_c$=7.2K was found in this metal. The quasiparticle contribution
to the Nernst signal attains a magnitude comparable to the vortex
signal in the superconducting state. Comparing the evolution of
Nernst and Hall coefficients, we observed that the maximum in
Nernst signal occurs when the contribution of hole-like and
electron-like carriers to the Hall conductivity cancel out.
Moreover, we found that in the Charge Density Wave(CDW) state,
Nernst coefficient becomes sublinear as a function of magnetic
field. Our study recalls that the ambipolar flow of quasiparticles
in presence of a thermal gradient can lead to an enhancement of
the Nernst signal in a multi-band metal. The results open a new
window on the driving mechanism of the CDW instability in NbSe$_2$
and establish that a large sublinear Nernst signal can arise in a
metal in total absence of superconducting fluctuations.

Single crystals of 2H-NbSe$_{2}$ were grown by standard iodine
vapor transport method. Stoichiometric amounts of 99.9 percent
pure Nb wire and 99.999 percent pure Se shots were sealed in a
quartz ampoule, and then heated in a temperature gradient for a
few weeks. On each sample, four longitudinal and two lateral
electrodes were painted with silver epoxy in order to measure the
resistivity($\rho_{xx}$) and the Hall coefficient (R$_H$). Nernst
coefficient, thermopower and thermal conductivity were measured
using a one-heater-two-thermometer set-up which allowed us to
measure all thermoelectric coefficients of the sample in the same
conditions.

\begin{figure}
\resizebox{!}{0.35\textwidth}{\includegraphics{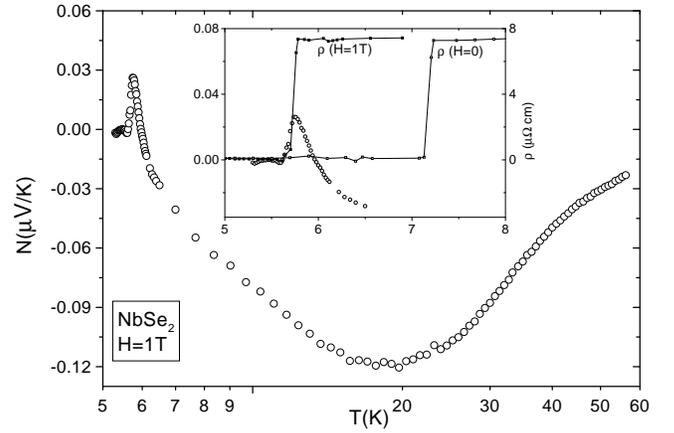}}
  \caption{\label{fig1}Nernst effect in NbSe$_2$ in a semi-logarithmic plot.
 A sharp positive signal associated with thermally-induced vortex
 movement is superposed on a large negative quasi-particle signal. The
insert compares  the behavior of Nernst coefficient and
resistivity near the superconducting transition.}
\end{figure}

The temperature-dependence of the Nernst coefficient in NbSe$_2$
at H=1T is displayed in Fig.1. A sharp peak associated with the
superconducting transition is superposed on a large negative
signal which peaks at 20K. As seen in the inset, the positive peak
occurs in a temperature window closely related to the
superconducting transition. The sharp resistive transition
($\Delta$T$_c\sim$0.1K) at H=1T indicates that the vortex liquid
state occurs only in a very narrow temperature-field window
stretching along the H$_{c2}$(T) line in the (H,T) plane. Since
the vortex Nernst signal changes drastically in a very narrow
temperature interval, the presence of a relatively large
temperature gradient along the sample is expected to broaden the
peak. No systematic study of the variation of the signal with the
magnitude of the temperature gradient was performed. The size of
the peak (0.03 $\mu$V/K), observed in presence of a temperature
difference of about 0.2K between the middle electrodes,
underestimates the magnitude of the maximum vortex signal.

One striking feature of Fig.1 and the main new result of this
investigation is the presence of a large negative Nernst signal in
the normal state which presents a broad maximum around 20K. Like
many other two-dimensional dichalcogenides, NbSe$_2$ undergoes a
CDW transition at T$_{CDW}\sim$ 32K\cite{moncton}. In order to
explore a possible connection between the anomalously large Nernst
signal and the CDW transition, we measured the temperature
dependence of thermal conductivity ($\kappa$), thermopower (S) and
Hall coefficient of the same sample.

 Fig.2 displays the temperature dependence of longitudinal
(thermal, electric and thermoelectric) conductivities.
Thermopower, slightly increasing with decreasing temperature for
temperatures above T$_{CDW}$, presents a broad maximum and then,
at temperatures well below T$_{CDW}$, displays a purely linear
temperature dependence. This linear decrease is only interrupted
with the superconducting transition. The application of a magnetic
field of 5T, strong enough to destroy superconductivity, restores
the linear S(T) without any detectable field-induced change in the
magnitude of S. As seen in the lower panel, the effect of CDW
transition on charge and heat transport is far from spectacular.
As reported in previous studies \cite{lee,harper,corcoran}the
resistivity presents a barely noticeable anomaly at T$_{CDW}$. We
observed a slight gradual enhancement of charge
conductivity,$\sigma$  below T$_{CDW}$. A concomitant enhancement
is also observable in the temperature dependence of $\kappa$/T. A
rough estimate of the size of the electronic contribution to heat
transport can be obtained by multiplying $\sigma$ by the
Sommerfeld number (L$_0$=24.5 10$^{-9} W\Omega/K^2$), and
comparing it with $\kappa$/T. As seen in the figure, the fraction
of heat conductivity due to electrons increases from thirty
percent at 35 K to eighty percent at the onset of
superconductivity.

\begin{figure}
\resizebox{!}{0.7\textwidth}{\includegraphics{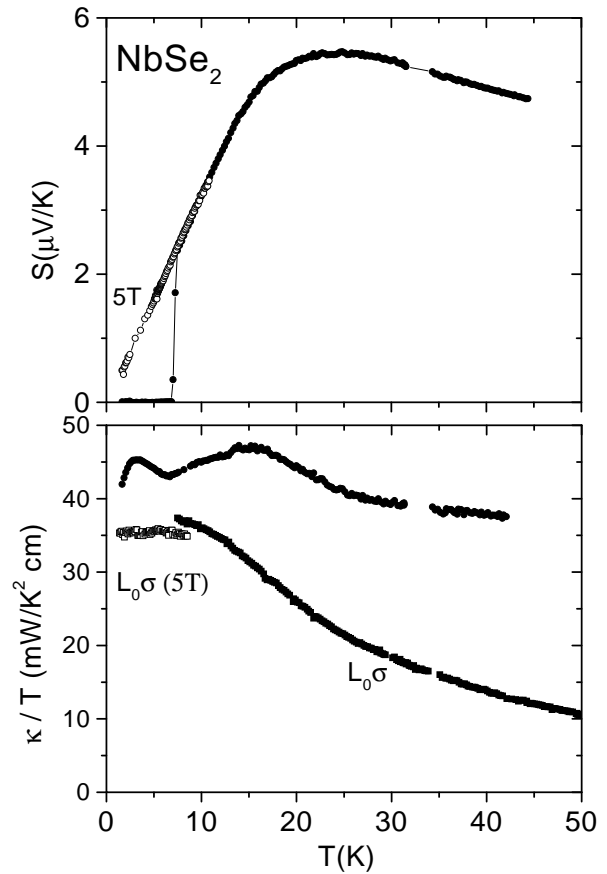}}
  \caption{\label{fig2} Upper panel: thermopower(S) of NbSe$_2$
at H=0 (solid circles) and H=5T (open circles). Lower panel:
Thermal conductivity divides by temperature (solid circles) as a
function of temperature. Also shown is the charge
conductivity($\sigma$) at H=0 (solid squares) and at H=5T(open
squares) multiplied by the constant L$_0$ (see text). }
\end{figure}

The diagonal thermoelectric coefficients, R$_H$ and N, plotted in
Fig.3, present more remarkable signatures of the CDW transition
which contrasts with the behavior observed for $\sigma$ and
$\kappa$/T. Upon cooling from room
 temperature, the Hall coefficient remains virtually constant\cite{lee}. At T=32K,
 with the entry of the system into the CDW state, it begins to deviate downward
 from this positive constant value (+4.9 10$^{-10}$ m$^{3}$/C). The decrease of the Hall
 coefficient continues down to 8K and then saturates to a constant negative
 value (-6.1 10$^{-10}$ m$^{3}$/C). One can see in the upper panel of Fig.3,
 that the Nernst coefficient, which remains negative above T$_c$ in the
whole temperature window investigated, presents a peak at T=21K.
Remarkably, at this temperature R$_H$ is almost zero. We will
argue below that this gives an important clue to the origin of the
Nernst signal. A second feature of data is revealed by comparing
the field-normalized Nernst signal (N/H) at two different fields
(H=1T and H=5T). The two curves superpose for T $>$ 27K but become
clearly distinct for lower temperatures. This suggests that the
Nernst signal ceases to be  field-linear in the CDW state. As seen
in the insert which compares the field dependence of N at three
different temperatures, N(H) which is linear at 34K, becomes
clearly sublinear at 16K. Also shown in the figure is the T=4K
curve. Here N is virtually zero up to H$\simeq$ 2.3T, then it
becomes positive in a narrow field associated with vortex movement
before attaining the normal-state negative regime.

\begin{figure}
\resizebox{!}{0.7\textwidth}{\includegraphics{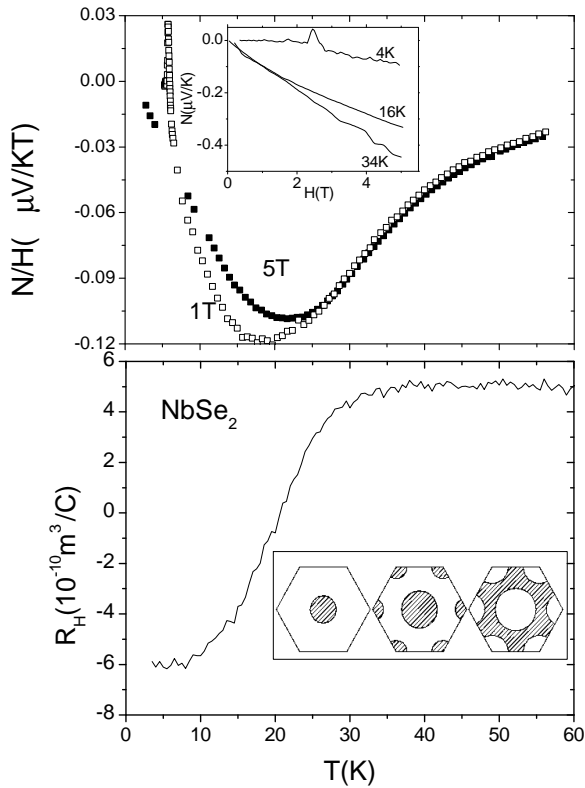}}
  \caption{\label{fig3}Upper panel: Nernst coefficient divided by magnetic field
as a function of temperature at H=0 and H=5T. The insert compares
the field-dependence of the Nernst coefficient at three different
temperatures. Lower panel: The temperature-dependence of the Hall
coefficient measured at H=5T. Insert: A schematic plot of the
three-band Fermi surface in NBSe$_2$ as observed by
ARPES\cite{yokoya}.}
\end{figure}

Band calculations\cite{corcoran,rossnagel} have predicted a
complex Fermi Surface(FS) for NbSe$_2$ which consists of a small
hole-like closed pocket, two hole-like cylinders and two
electron-like cylinders (see the insert in the lower panel of Fig.
3). While only the small hole-like pocket was detected by de
Hass-van Alphen measurements\cite{corcoran}, more recent
Angular-Resolved Photoemission Spectroscopy (ARPES) studies have
led to the detection of all portions of the predicted FS
\cite{yokoya,rossnagel}.

Now, in presence of such a complicated FS, a finite Nernst signal
is not unexpected. Following Wang \emph{et al.}\cite{wang1}, we
define the Peltier conductivity tensor $\overline{\alpha}$ with
equations $\overrightarrow{J}=\overline{\sigma} \overrightarrow{E}
- \overline{\alpha }\overrightarrow{\nabla}T$ and
$\overrightarrow{J_q}=\overline{\alpha} T \overrightarrow{E} -
\overline{\kappa}\overrightarrow{\nabla}T$. Here,
$\overrightarrow{J}$ and $\overrightarrow{J_q}$ are charge and
heat current densities. $\overrightarrow{E}$ and $\nabla T$ are is
the electric field and the thermal gradient. $\overline{\sigma}$
and $\overline{\kappa}$ are electric and thermal conductivity
tensors. Assuming $\sigma_{xy}\ll \sigma_{xx}$ and neglecting the
transverse thermal gradient produced by a finite $\kappa_{xy}$,
the Nernst coefficient is  easily obtained as\cite{wang1}:

\begin{eqnarray}
  N &=& \frac{E_y}{\frac{\partial T}{\partial x}} =
  S(\frac{\alpha_{xy}}{\alpha_{xx}}-\frac{\sigma_{xy}}{\sigma_{xx}})
\end{eqnarray}

Where $S=\frac{\alpha_{xx}}{\sigma_{xx}}$ is the thermopower. For
a single band, and if $\overline{\sigma}$ is not energy-dependent,
one has:
\begin{eqnarray}
\frac{\sigma_{xy}}{\sigma_{xx}} &=&
\frac{\alpha_{xy}}{\alpha_{xx}}
\end{eqnarray}

and the two terms in Eq.1 cancel out (``Sondheimer
cancellation'')\cite{wang1}. Now, let us assume that the metal is
not single band and there are two FS sheets with dominant carriers
of opposite signs. Then Eq.1 becomes:
\begin{eqnarray}
N  &=& S(\frac{\alpha^{+}_{xy}+\alpha^{-}_{xy}}
{\alpha^{+}_{xx}+\alpha^{-}_{xx}}-\frac{\sigma^{+}_{xy}+\sigma^{-}_{xy}}{\sigma^{+}_{xx}+\sigma^{-}_{xx}})
\end{eqnarray}
\\
Where the superscript designates the sign of the dominant
carriers. Now, obviously, the validity of Eq.2 for each band does
\emph{not} lead to the cancellation of the two terms in the right
side of Eq.3.  We can readily see that in a compensated two-band
system, i.e. in the case of $\sigma^{-}_{xy}$=-$\sigma^{+}_{xy}$,
the second term on the right side of Eq.3 vanishes. But, since
$\alpha^{-}_{xx}$ and $\alpha^{+}_{xx}$ are expected to have
different signs, Eq.2 implies the same sign for
$\alpha^{\pm}_{xy}$. Therefore, the first term does not vanish and
would yield a finite Nernst signal.

As recalled above, NbSe$_{2}$ is a multi-band metal and becomes
compensated at T=21K. Therefore, the finite size and the
temperature dependence of the Nernst signal found in our study can
safely be attributed to the counterflow of carriers with opposite
sign. In semiconductors, this phenomenon, known  as the ambipolar
Nernst effect, has been known since a long time ago\cite{delves}.
However, to our knowledge, this is the first case of a metal
displaying the effect.

Using the experimental data and Eq.1, one can compute the
temperature dependence of the two components of the Peltier
conductivity tensor $\overline{\alpha}$\cite{approx}. Fig. 4
compares the two ratios $\frac{\sigma_{xy}}{\sigma_{xx}}$ and
$\frac{\alpha_{xy}}{\alpha_{xx}}$.  As seen in the figure, at the
onset of the CDW transition, the two angles display opposite signs
and the absolute magnitude of $\frac{\alpha_{xy}}{\alpha_{xx}}$ is
five times larger than $\frac{\sigma_{xy}}{\sigma_{xx}}$. Below
27K, the two angles begin to gradually converge to a comparable
negative magnitude. Now, $\overline{\alpha} \propto
\frac{\partial\overline{\sigma}}{\partial\epsilon}|_{\epsilon=E_{f}}$,
for each band. Therefore, the contrasting behavior observed here
indicates that the energy dependence of $\sigma_{xy}$ and
$\sigma_{xx}$ is substantially different on various bands.

These results provide fresh input for the ongoing effort to
identify the driving mechanism of the CDW instability in
NbSe$_{2}$. Surprisingly, even the recent high-resolution ARPES
study which successfully probed the anisotropy of the
superconducting gap\cite{yokoya} failed to detect a CDW gap.
Moreover, the slightly incommensurate CDW vector, revealed by
neutron scattering \cite{moncton}, can not be
 associated in any obvious way with a nesting vector of the known FS
\cite{straub,rossnagel}. The observation of a FS in good agreement
with the theoretically-predicted one as well as the absence of any
detectable gap indicates that the CDW transition is not
accompanied by any substantial modification of the Fermi
surface\cite{rossnagel}. The temperature dependence of specific
heat close to the CDW instability\cite{harper} confirms that the
change in the density of states is small. The above-mentioned
behavior of $\sigma(T)$ and $\kappa(T)$ point to the same
conclusion.

But, if the CDW transition leaves the FS almost intact, how to
account for the spectacular sign change of the Hall coefficient?
One plausible scenario would be a drastic change in electronic
mean-free-path induced by the transition. To illustrate the point,
let us use Ong's geometrical picture of two-dimensional Hall
conductivity\cite{ong}. In a metal with a hole-like and
an-electron FS with circular cross-sections, one can write:
\begin{eqnarray}
R_{H}&=&\frac{2\pi d (\ell^{2}_{+}-\ell^{2}_{-})}{e
[(k^{+}_{F}\ell_{+})^{2}+(k^{-}_{F}\ell_{-})^{2}]}
\end{eqnarray}

where k$^{\pm}_{F}$  and $\ell_{\pm}$ are Fermi vector and
mean-free-path for electrons and holes. Here, the latter is
assumed to be isotropic for each band (the ``isotropic-l''
approximation). A drastic increase in $\ell_{-}$ below T$_{CDW}$
leads to a change of sign of R$_{H}$ without any modification in
the FS. The case for an unusually high mean-free-path  for the
electron-like orbit is supported by another piece of experimental
evidence which is the contrasting effect of the impurities on
$\rho_{xy}$ and $\rho_{xx}$. Improvement in sample quality leads
to a small decrease of residual resistivity, but a much larger
enhancement of the negative Hall signal at low
temperatures\cite{huntley,jing}.

The emergence of sublinearity in the field dependence of the
Nernst coefficient can also be explained in this scenario. [A
similar sublinearity is reported for the Hall coefficient in clean
samples\cite{huntley}. We did not check the field-dependence of
the Hall coefficient in our study.] It indicates a negative
coefficient for the H$^3$ term in the Zener-Jones expansion and
can be related to a small electron-like portion of the FS with a
long mean free path\cite{mackenzie}.

Thus, it is tempting to conclude that the CDW transition is
accompanied with a sharp change in the scattering rate on an
electron-like orbit. We note that a drastic change in scattering
rate is naturally expected in the model proposed by Rice and Scott
in which the existence of saddle points close to the Fermi surface
drives the formation of the CDW\cite{rice}. While these saddle
points have been detected by ARPES measurements, their separation
in the k-space does not correspond to the CDW
 vector\cite{straub,rossnagel}. It would be interesting to explore
 the consequences of this scenario for the thermoelectric
 coefficients.

This study presents an example of various possible origins for a
large sublinear Nernst signal in a metal. It provides an
interesting information for the debate on the origin of the Nernst
signal observed in cuprates. We recall that in electron-doped
cuprates, the quasi-particle contribution to the Nernst signal is
large and field-linear and can be easily distinguished from the
vortex contribution. The large magnitude of the latter (the signal
attains a maximum of  0.1-0.2 $\mu V/ KT$, close to the value
found here for NbSe$_2$), has been attributed to the existence of
a two-band FS\cite{fournier,gollnik}. In the hole-doped cuprates,
a smaller sublinear Nernst signal is present in temperatures well
above T$_c$ and can not be distinguished from the vortex
signal\cite{xu,wang1,wang2,capan}. While strong superconducting
fluctuations would provide a natural explanation for this signal,
one should not disregard alternative scenarios connected with
subtle changes in the electronic properties of the normal state.

\begin{figure}
\resizebox{!}{0.35\textwidth}{\includegraphics{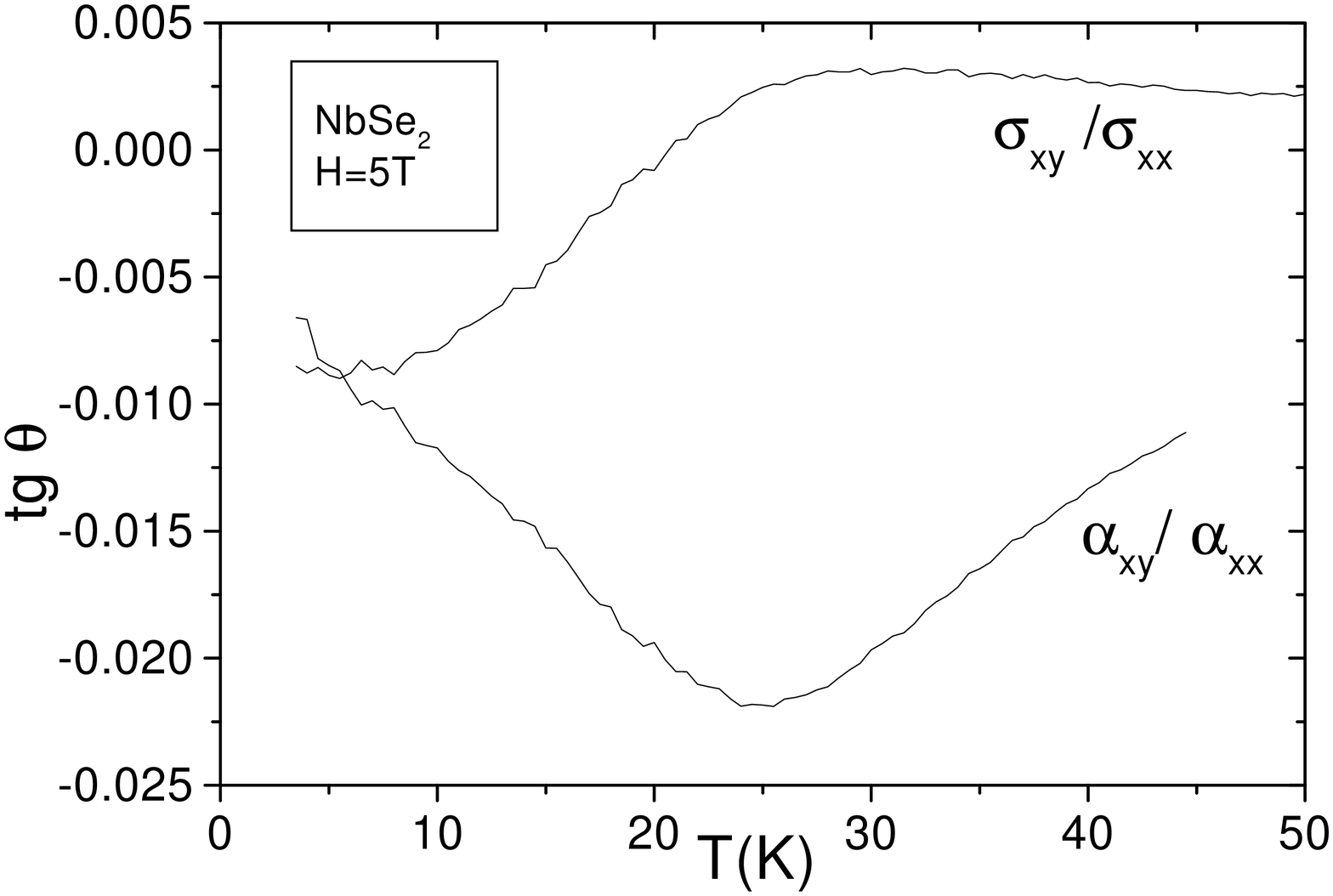}}
  \caption{\label{fig4} The temperature dependence
of two Hall-like angles (see text).}
\end{figure}

In conclusion, we found that the ambipolar flow of quasi-particles
large quasi-particle leads to a large contribution to Nernst
coefficient in NbSe$_2$. The behavior of the off-diagonal
components of two conductivity tensors indicate a drastic change
in electron scattering induced by the CDW transition.

We thank T. H. Geballe for fruitful discussions and H. Richter for
technical assistance.


\begin{thebibliography}{99}
\bibitem{approx} Note that this procedure neglects the transverse thermal gradient which appears in our adiabatic set-up
in presence of a finite $\kappa_{xy}$. The correction is expected
to be small at 30K and to increase with decreasing temperatures as
the relative weight of the electronic thermal conductivity
gradually enhances.
\end{thebibliography}
\end{document}